\documentclass[pra,10pt,twocolumn]{revtex4-1}
\pdfoutput=1
\usepackage{graphicx,amsmath,amssymb,upgreek}
\usepackage[utf8x]{inputenc}
\usepackage{hyperref}
\usepackage{appendix}

\begin{document}

\title{Walk dimension for light in complex disordered media}

\author{Romolo Savo$^{1,2}$}
\email{romolo.savo@lkb.ens.fr}
\altaffiliation{Current affiliation: Laboratoire Kastler Brossel, UMR 8552, CNRS, Ecole Normale Sup\'erieure, Universit\'e Pierre et Marie Curie, Coll\`ege de France, 24 rue Lhomond, 75005 Paris, France}
	
\author{Matteo Burresi$^{1,3}$}
\author{Tomas Svensson$^{1}$}
\author{Kevin Vynck$^{1,4}$}
	\altaffiliation{Current affiliation: Laboratoire Photonique, Num\'erique et Nanosciences (LP2N), UMR 5298, CNRS - IOGS - Univ. Bordeaux, Institut d'Optique d'Aquitaine, 33400 Talence, France}
\author{Diederik S. Wiersma$^{1,2,3}$}

\affiliation{$^1$European Laboratory for Non-linear Spectroscopy (LENS), 50019 Sesto Fiorentino (FI), Italy}
\affiliation{$^2$Universit\`a di Firenze, Dipartimento di Fisica e Astronomia, 50019 Sesto Fiorentino (FI), Italy}
\affiliation{$^3$Istituto Nazionale di Ottica (CNR-INO), Largo Fermi 6, 50125 Firenze (FI), Italy}
\affiliation{$^4$Institut Langevin, ESPCI ParisTech, CNRS, 1 rue Jussieu, 75238 Paris Cedex 05, France}


\begin{abstract}
Transport in complex systems is characterized by a fractal dimension -- the walk dimension -- that indicates the diffusive or anomalous nature of the underlying random walk process. Here we report on the experimental retrieval of this key quantity, using light waves propagating in disordered media. The approach is based on measurements of the time-resolved transmission, in particular on how the lifetime scales with sample size.  We show that this allows one to retrieve the walk dimension and apply the concept to samples with varying degree of fractal heterogeneity. In addition, the method provides the first experimental demonstration of anomalous light dynamics in a random medium. 
\end{abstract}

\maketitle

\section{Introduction}
Despite its conceptual simplicity a random walk can describe several and very different transport phenomena. Examples include heat transport~\cite{reif2009fundamentals}, animal foraging~\cite{benichou2011intermittent}, human travel~\cite{Brockmann2006_Nature}, molecular diffusion~\cite{barkai2012_PhysToday}, 
and the propagation of waves in random media~\cite{Ishimaru1999_Book, marshak20053d}. 
A fundamental aspect of random walks in disordered media is the scaling property of the dynamics, which is commonly expressed by the following type of propagator~\cite{benAvrahamHavlinBook2000,Cates1985}:
\begin{equation}
W(\mathbf{r},t|\mathbf{r'}) =  t^{d_\mathrm{w}/d_\mathrm{f}} \tilde{W} \left(\frac{|\mathbf{r}-\mathbf{r'}|}{t^{1/d_\mathrm{w}}}\right),
\label{eq:propagator}
\end{equation}
where $\tilde{W}$ is an invariant profile function, $d_\mathrm{f}$ is the (mass) fractal dimension of the system and $d_\mathrm{w}$ is the walk dimension (namely the fractal dimension of the random walk). This relation expresses the general fact that the propagator $W$, which provides the probability density to find a random walker in point $\mathbf{r}$ at time $t$ starting from $\mathbf{r}'$ at $t=0$, can be rescaled onto $\tilde{W}$ at any time $t$. Most importantly, the propagator $W$ is expected to spread asymptotically with the characteristic length $\xi(t)\sim t^{1/d_\mathrm{w}}$. 
As the value of the walk dimension reveals unambiguously whether the process is diffusive ($d_\mathrm{w}=2$) , subdiffusive ($d_\mathrm{w}>2$) or superdiffusive ($d_\mathrm{w}<2$), it has become a key observable in the anomalous transport research~\cite{benAvrahamHavlinBook2000, Klages2008_book,Condamin2007_Nature}.
Measurements have been performed on the random walk of small objects, like particles and molecules~\cite{Li2006_PRE, berkowitz2000anomalous, palombo2011spatio,tolic2004anomalous, golding2006_PRL, Szymanski2009_PRL,barkai2012_PhysToday} . In these cases,  the actual trajectories of the objects could be monitored in time \emph{inside} the complex medium and thus Eq.~(\ref{eq:propagator}) could be directly tested. Similarly to massive objects, multiply scattered waves in disordered media may also propagate anomalously~\cite{Davis1997,marshak20053d},
in which case the experimental retrieval of $d_\mathrm{w}$ is particularly challenging.  In recent years superdiffusion of light has been investigated in disordered photonics materials dubbed L\'evy glasses~\cite{Barthelemy2008_Nature,Bertolotti2010_AFM} and in hot atomic vapors~\cite{Mercadier2009_NatPhys, baudouin2014_arxiv}.  However, neither experiments on the dynamics of anomalous transport have been performed nor the walk dimension $d_\mathrm{w}$ has been measured. 

Here we report on the experimental retrieval of the fractal dimension of an optical random walk, by analyzing the scaling of the time-resolved transmission with sample size. We show that this type of analysis can clearly distinguish between different transport 
regimes -  from diffusive to superdiffusive -  consistent with the topology of the system.  Apart from the retrieval of $d_\mathrm{w} $ for an optical random system, this study also provides a direct confirmation of superdiffusion in L\'evy glasses. The concept is general  and applies to other transport phenomena. 

\section{Samples and experiment}
L\'evy glasses provide a convenient playground for the investigation of anomalous diffusion since their structural parameters can be readily varied. Moreover, the use of light waves has the enormous advantage that transport can be studied via simple observables, and a large statistics can be accumulated. The L\'evy glasses used in these experiments are obtained by embedding glass microspheres with diameter $\o$ distributed with a power-law, $p(\o) \sim \o^{-(\beta+1)}$, into an index-matched polymer matrix containing randomly dispersed TiO$_2$ nanoparticles~(See Appendix~\ref{appendixA}). Light is scattered by the nanoparticles in between the microspheres, while propagating freely inside them. The power-law heterogeneity size distribution $p(\o)$ is created to yield a heavy-tailed step length distribution for the random walk of light, and is expected to lead to superdiffusion~\cite{Barthelemy2008_Nature, Bertolotti2010_AFM}. The ``degree of superdiffusivity'' may then be controlled via the exponent $\beta$ at the sample preparation stage. The structure is contained between two glass slides, index-matched with the polymer matrix and the microspheres, with the sample thickness $L$ approximately equal to the largest sphere diameter.

Measurements were also performed on samples containing an homogeneous random distribution of nanoparticles. These were obtained by substituting the microspheres of L\'evy glasses with an equivalent volume of polymer. Such samples are expected to be diffusive and differ from L\'evy glasses only in the spatial distribution of scatterers. In addition, given the equal volume-averaged density of nanoparticles, it is expected that the step length distribution of these samples differs from L\'evy glasses only in its second moment but not in its first moment (i.e., equal mean step)~\cite{svensson2013holey}.

The experiment was performed by shining 130~fs laser pulses at 810 nm wavelength on the sample and collecting the transmitted light. By adopting a nonlinear optical gating technique~\cite{Shah:1988kl,svensson2013exploiting,Wiersma:1999kl} we monitor the temporal profile of the transmission with sub-ps resolution [Fig.~\ref{fig:1}(a)] (see Appendix~\ref{appendixB}).

\section{Results and Discussion}
\subsection{Time-resolved transmission}

\begin{figure}
\begin{center}
\includegraphics[]{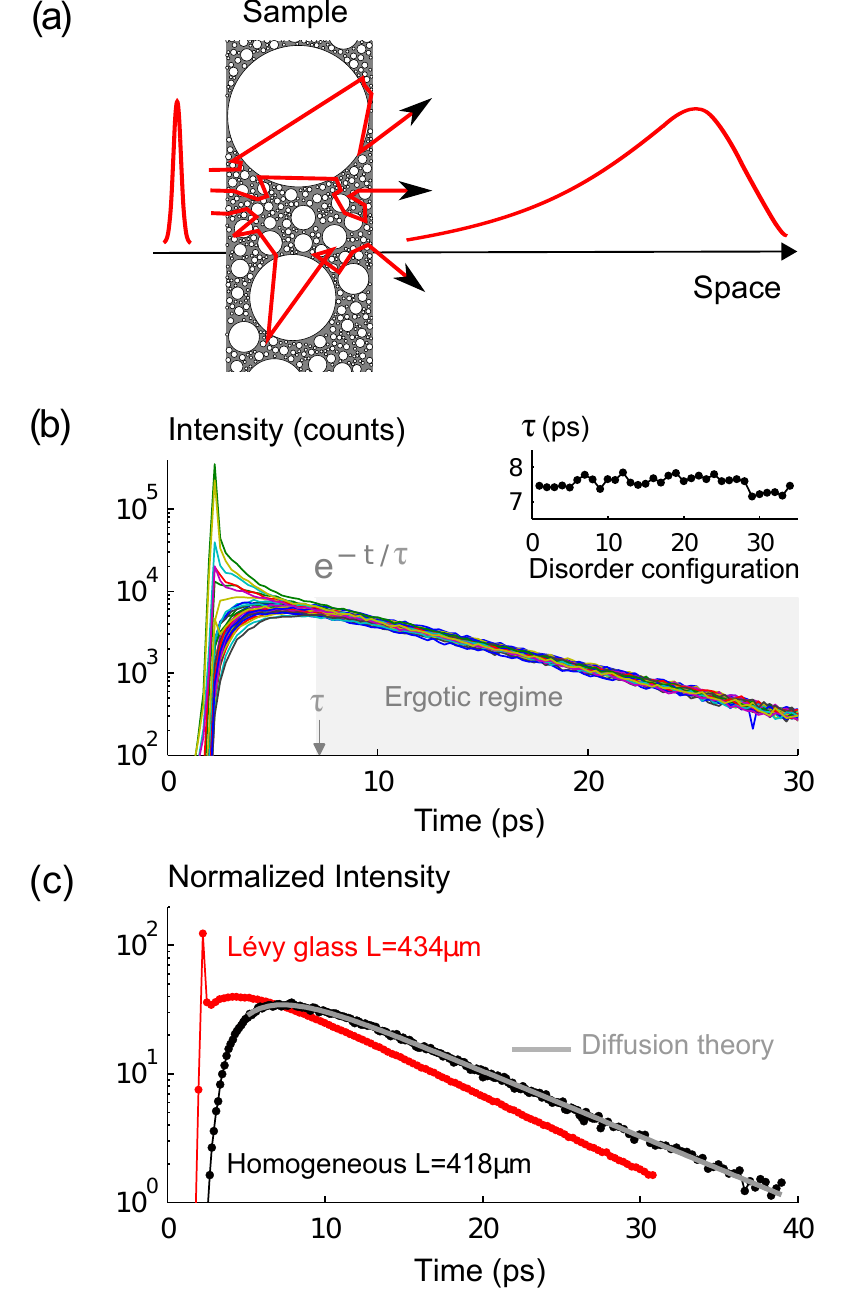}
\end{center}
\caption{(Color online) Time-resolved transmission. (a) Sketch of the experimental setting. A sub-ps laser pulse is incident onto a L\'evy glass and the time-resolved transmission is measured. (b) Measured time-resolved transmission for a L\'evy glass with $\beta=2.0$ and $L=434~\upmu$m for different injection points on the surface. The early-time response strongly depends on the source position while at late times, the intensity decays exponentially with a narrow lifetime distribution (see the inset). (c) Average time-resolved transmission for the L\'evy glass in (b) and a homogeneous sample. The two samples have a comparable thickness and the same volume-averaged density of TiO$_2$ nanoparticles. The curves have been normalized to the laser pulse power. The time-resolved transmission responses are strikingly different, indeed the first light arrives later and the lifetime is longer for the homogeneous sample. A fit with diffusion theory returns a transport mean free path $\ell_t=40~\upmu$m.}
\label{fig:1}
\end{figure}

Figure~\ref{fig:1}(b) shows a series of measurements of the time-resolved transmission performed on a L\'evy glass ($\beta=2.0$, $L=434 \upmu$m) for different injection points on the sample surface. The strong dependence of the early-time profile with the injection point is evident, in sharp contrast with the late-time behavior, where an exponential tail $\exp(-t/\tau)$ with a well-defined lifetime $\tau$ is observed (see the inset). The intensity fluctuation at early times is transient, and due to the large-scale heterogeneities 
of the sample \cite{Bertolotti2010_PRL, Burresi2012_PRL} which allow light to escape from deep inside in only a few scattering events. 
The fluctuations vanish at times larger than $\tau$, which is approximately the time needed to explore all regions of the sample (the 
Thouless time~\cite{Akkermans2007_Book}) and reach ergodicity. This transient, position-dependent, dynamics is not observed in statistically homogeneous systems.

Figure~\ref{fig:1}(c) shows the comparison between the time-resolved transmission of the L\'evy glass and the reference homogeneous sample. The average profile for the homogeneous sample does not exhibit any sharp features at early times, as expected. 
A fit with diffusion theory~\footnote{The early-time transmission cannot be fitted due to a transient failure of diffusion theory in thin slabs ($L/\ell_t<$10)~\cite{elaloufi2004_JOSA}. Hence the fit is taken over a range that starts around the maximum and covers the entire tail. This corresponds to the range over which the grey line is drawn.}  returns a transport mean free path $\ell_t=40~\upmu$m, which we shall take as the mean step length $\langle \ell \rangle$ in our L\'evy glasses.

\subsection{Lifetime scaling with sample thickness}
\begin{figure} [t!]
\begin{center}
\includegraphics[]{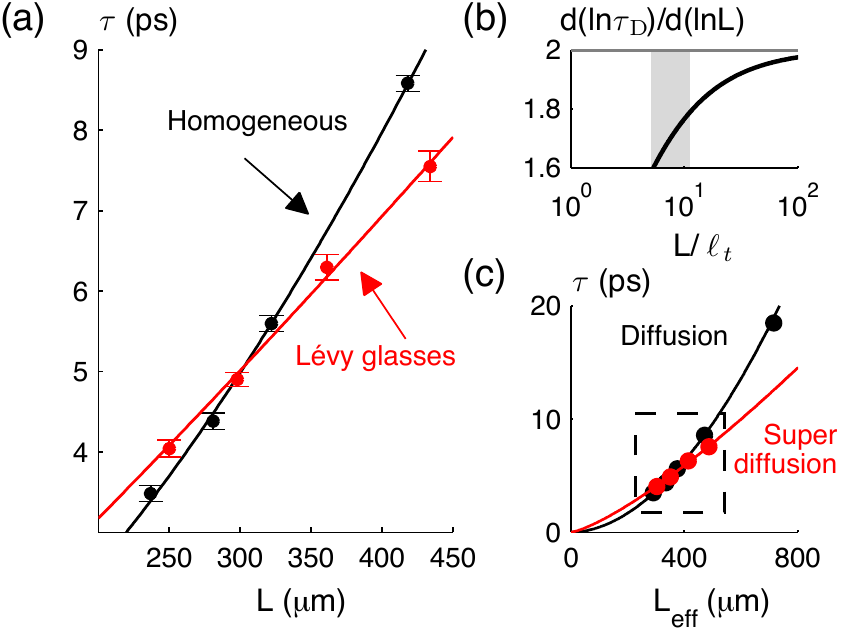}
\end{center}
\caption{(Color online) Scaling of lifetimes. (a) Lifetimes versus sample thickness for the $\beta=2.0$ L\'evy glass and the homogenous media. Error bars give the standard deviations of measurements performed in different injection points. Markedly different scaling trends are found. The scaling exponent $z$ is obtained from the fit $\tau=bL^z$ (continuous lines). (b) Calculation of $\mathrm{d}(\mathrm{ln}(\tau_{d}))/\mathrm{d}(\mathrm{ln}(L))$ versus $L$ from diffusion theory shows the effect of finite system size on the evaluation of $z$. The highlighted interval corresponds to the thickness range of investigation which returns a value in good agreement with the measured one. (c) The scaling exponent $d_\mathrm{w}$ is retrieved from the fit $\tau=cL_\mathrm{eff}^{d_\mathrm{w}}$. The additional reference data point shows that at larger $L$ the deviation between diffusive and superdiffusive scaling is expected to be even clearer. Unfortunately, so far, we could not realize reliable L\'evy glasses at larger thicknesses. The fits for the two data sets are performed on the same thickness range.}
\label{fig:2}
\end{figure}

As shown in Ref.~\cite{Buonsante2011_PRE}, the time-resolved transmission is expected to verify asymptotically the following scaling behavior:
\begin{equation}
T(t,L)=L^\gamma \tilde{T} \left (\frac{L}{\xi(t)} \right),
\label{eq:scalingtransmission}
\end{equation}
where $\tilde{T}$ is an invariant transmission function and $\gamma$ a characteristic exponent related to the steady-state transmission properties. Furthermore, given the exponential tail of the transmission curve $T(t) \sim \exp(-t/\tau)$ (see Fig.~\ref{fig:1}), we expect the lifetime to scale as $\tau \sim L^{d_\mathrm{w}}$.
Hence, we performed time-resolved transmission experiments on samples with different thicknesses to retrieve the scaling exponent $d_\mathrm{w}$.  We considered L\'evy glasses with two heterogeneity size distributions ($\beta=2.0$ and $2.6$) and the corresponding homogeneous samples. The volume-averaged density of nanoparticles was kept constant in all samples for the sake of comparison. The importance of this point will be discussed later on. The lifetime $\tau$ was obtained from the decay at long times, using an average over 30 different injection points, leading to a standard deviation smaller than $3\%$.

As we are considering real (finite) systems a quantitative analysis of their scaling properties requires to account for the presence of boundaries. However, we prefer to start by showing the raw data since they already allow for important physical considerations. Figure~\ref{fig:2}(a) shows the measured $\tau$ as a function of $L$ for L\'evy glasses with $\beta=2.0$ and the homogeneous samples. Their dependence with sample thickness turns out to be very different, as a consequence of a pure spatial redistribution of the scatterers. We quantify this difference by performing a fit to the data with a generic power-law function $\tau = b L^z$, where $b$ is a constant and $z$ is the measured scaling exponent. The fits return $z=1.1\pm0.2$ for the L\'evy glass and $z=1.6\pm0.2$ for the homogeneous media, thereby revealing a remarkably different dynamics. The smaller $z$ in the L\'evy glass indicates a faster growth of the characteristic length $\xi$ with time, originating from the heavy-tailed step length distribution. The same analysis performed on the $\beta=2.6$ L\'evy glass yields $z=1.3\pm0.2$. This shows that the light transport dynamics is dictated by the heterogeneity size distribution, via the $\beta$ exponent.

We now consider the effect of system finiteness on the evaluation of the scaling exponent. It is well known that the diffuse lifetime in finite systems should scale as $\tau_D \propto (L+2z_e)^2$, where $z_e$ is the so-called ``extrapolation length''~\cite{Ishimaru1999_Book}. 
Since $z_e=\frac{2}{3}\ell_t$ for three-dimensional systems with index matched boundaries 
(neglecting internal reflections)~\cite{vanrossum1999_book}, when $L \gg \ell_t$ the contribution of $z_e$ to $L_e$ is negligible ($L_e\simeq L$) and thus $z=d_\mathrm{w}=2$. However, for our homogeneous sample with $\ell_t = 40~\upmu$m and thicknesses in the range $230 \leq L \leq 450~\upmu$m, we expect a difference between $z$ and $d_\mathrm{w}$. This is evident also from Fig.~\ref{fig:2}(b), where $z=\mathrm{d}(\mathrm{ln}\tau_{D})/\mathrm{d}(\mathrm{ln}L)$ is reported as a function of the optical thickness $L/\ell_t$. The exponent $z$ obtained in the thickness range under consideration is in good agreement with the experimentally retrieved value ($z=1.6\pm0.2$) and the limit $z \rightarrow d_\mathrm{w}=2$ is reached only for thicker samples. 

Since there is currently no theory available to calculate the extrapolation length for L\'evy glasses, we use the diffusive case as a starting point and assume that the relation $z_e=\frac{2}{3}\ell_t=\frac{2}{3}\langle \ell \rangle$ provides at least the right order of magnitude of the extrapolation
length~\footnote{The precise determination of the extrapolation length is complicated even in regular diffusive samples and is influenced e.g. by internal reflection. Recent elaborate numerical studies, which will be published elsewhere, indicate that indeed $z_e$ remains of the order of $\ell $}.  Both homogeneous samples and L\'evy glasses were prepared to have the same mean step length~\cite{svensson2013holey}, which simplifies the analysis.  Power-law fits of the lifetime data using $\tau = c L_\mathrm{eff}^{d_\mathrm{w}}$ are shown in Fig.~\ref{fig:2}(c) and the results are summarized in Table~\ref{tb:scaling exponents}. The scaling exponent $d_\mathrm{w}$ obtained for the homogeneous sample is fully consistent with diffusive transport and those for the L\'evy glasses are markedly smaller than 2. 

\begin{table}
\centering
\begin{tabular}{l c c}
Samples set & $z$ & $d_\mathrm{w}$ \\
\hline
L\'evy Glasses $\beta$=2.0 & 1.1$\pm$0.2 & 1.3$\pm$0.3 \\
L\'evy Glasses $\beta$=2.6 & 1.3$\pm$0.2 & 1.5$\pm$0.2 \\
Homogeneous samples & 1.6$\pm$0.2 & 1.9$\pm$0.2
\end{tabular}
\caption{Measured scaling exponents. The values reported for $z$ and $d_\mathrm{w}$ have been obtained as best fit exponents of the functions $\tau = b L^z$ and $\tau=cL_\mathrm{eff}^{d_\mathrm{w}}$, respectively. Errors have a statistical confidence of 95$\%$.}
\label{tb:scaling exponents}
\end{table}

\subsection{Data collapse}
\begin{figure} [t]
\begin{center}
\includegraphics{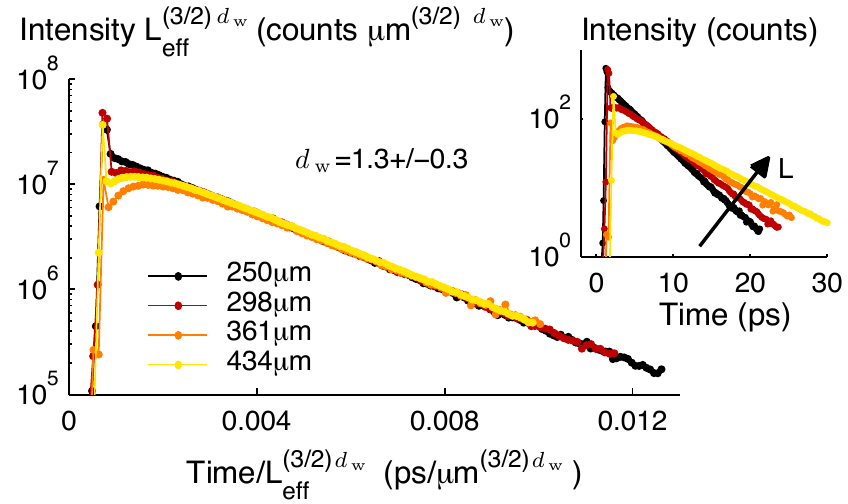}
\end{center}
\caption{(Color online) Collapse of the time-resolved transmission measured on L\'evy glasses. An appropriate rescaling of the axes, according to Eq.~(\ref{eq:scalingtransmission}) with the experimentally retrieved $d_\mathrm{w}$ and $\gamma=-\frac{3}{2}d_\mathrm{w}$, yields a collapse of the time-resolved transmission curves obtained from samples with different thicknesses onto a single invariant profile. In the inset data are plotted by using regular axes and the dependence of the long-time decay on the growing thickness is marked by the arrow. The collapse obtained by rescaling the axes shows that the (anomalous) transport dynamics is described by a unique scaling exponent $d_\mathrm{w}$.}
\label{fig:3}
\end{figure}

It is interesting to visualize and analyze the collapse of the time-resolved transmission curves expected from the scaling hypothesis in Eq.~(\ref{eq:scalingtransmission}) (with $L_\mathrm{eff}$ instead of $L$). Results are shown in Fig.~\ref{fig:3}. The time axis was rescaled as $t \rightarrow t/L_\mathrm{eff}^{d_\mathrm{w}}$. Regarding the transmission axis, we found that a very good collapse of the exponential tails \textit{for all systems investigated here} was obtained using $T\rightarrow TL_\mathrm{eff}^{3d_\mathrm{w}/2}$, i.e. by taking $\gamma=-\frac{3}{2}d_\mathrm{w}$ in Eq.~(\ref{eq:scalingtransmission}). Quite remarkably, this is the value expected for annealed L\'evy walks~(See Appendix~\ref{appendixC}).

The overall superposition of the transmission curves is excellent, showing that the measured scaling exponent $d_\mathrm{w}$ constitutes a unique parameter describing anomalous dynamics in these systems. The deviation at early times is also expected since the quasi-ballistic trajectories do not obey the scaling hypothesis that describes the spreading of the superdiffusive propagator. This is again typical of L\'evy walk dynamics~\cite{Drysdale1998_PRE, Zumofen1993}.

\section{Concluding remarks}
We conclude by discussing some interesting points for future investigation. 
First, the standard approach to deal with the finiteness of open disordered systems, via the extrapolation length, should be revisited to take into account non-exponential step length distributions. For power-law distributions, this problem is directly related to the treatment of boundaries in the presence of spatial nonlocality~\cite{Zoia:2007fk}.
Second, our scaling analysis has been performed on a restricted thickness range while it would be interesting to investigate, both experimentally and theoretically, the limit towards infinite system size~\cite{Groth2012_PRE, Burioni2014_PRE}.
With respect to this important issue, there will be a difference between the case of fixed filling fraction of heterogeneities, and that of maximal filling.  A recent numerical study on two-dimensional (2D) systems~\cite{Burioni2014_PRE} has shown that the fixed filling fraction case eventually leads to normal diffusion, while the maximal filled case leads to superdiffusion, provided that the system converges towards 100\% voids surrounded by point scatterers.
It is hence of fundamental importance to define properly how the limit to infinite system size is taken.
Third, the scaling exponents retrieved experimentally here do not fully match with those expected from a ``chord-length'' model, i.e. $d_\mathrm{w}=\beta-1$~\cite{Barthelemy2008_Nature, Bertolotti2010_AFM}. Indeed, this simplified model neglects the possibility of crossing multiple voids in one step and of having correlations between successive steps due to quenched disorder, which are both expected to affect the dynamics~\cite{Barthelemy2010_PRE, Groth2012_PRE, svensson2013holey, svensson2014_PRE, Burioni2014_PRE}, and which deserve further attention. 

In summary, we have presented a first experimental investigation of the walk dimension $d_\mathrm{w}$ for light waves in disordered photonic structures. The retrieval of $d_\mathrm{w}$ is possible via a scaling analysis of the lifetime, and allows one to distinguish clearly between regular and superdiffusive transport dynamics in L\'evy glasses. These results show how the photonics of disorder 
can be used in the research on anomalous transport~\cite{benAvrahamHavlinBook2000,Klages2008_book}, and that it can provide a new setting to investigate wave diffusion on fractal media~\cite{Akkermans2010_PRL,Fernandez-Marin2012_PRA}. 
It will be interesting to investigate how the fractal dimension plays a role in interference effects, like weak and 
strong localization~\cite{ping2006introduction}, where the dimensionality of the system can make a crucial difference 
between the occurrence of extended or localized states.

\section*{Acknowledgments}
We wish to acknowledge Raffaella Burioni, Alessandro Vezzani, Enrico Ubaldi, Sepideh Zakeri, Marco Grisi, and the entire 
\textit{Optics of Complex Systems} group at LENS for fruitful discussions,  Thomas Huisman and Lorenzo Pattelli for their help with sample preparation, and financial support from the European Research Council (FP7/2007-2013) ,  ERC grant agreement n° [291349]. T.S. acknowledges funding from The Swedish Research Council (Grant 2010-887), K.V.  from LABEX WIFI under references ANR-10-LABX-24 and ANR-10-IDEX-0001-02 PSL$^\star$.

\appendix

\section{Sample manufacturing}
\label{appendixA}
\subsection{L\'evy glasses}
\begin{figure} [t!]
\begin{center}
\includegraphics[width=\columnwidth]{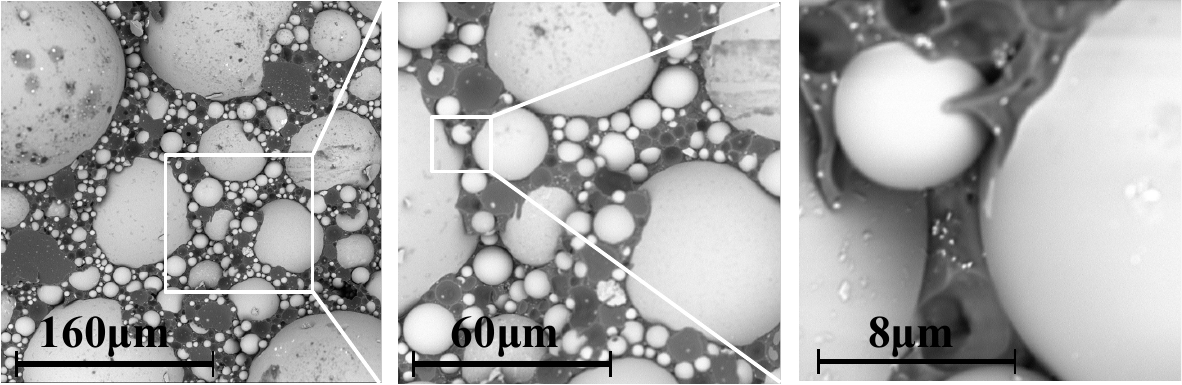}
\end{center}
\caption{Sequence of Scanning Electron Micrographs of the surface of a L\'evy glass with increasing magnification that reveals the statistical self-similarity of the heterogeneity. The hosting polymer appears in dark gray while the glass spheres (typically with $5 \leq \o \leq 400~\upmu$m)  in light gray. In the last close-up the TiO$_2$ nanoparticles are clearly distinguishable.}
\label{fig:4}
\end{figure}

L\'evy glasses (see Fig.~\ref{fig:4}) are obtained by first dispersing homogeneously the TiO$_2$ nanoparticles (Huntsman Tioxide R-XL, diameter 280~nm, refractive index 2.4) in a monomer host (acrylate optical glue Norlan 65, refractive index 1.52) and then by introducing Soda Lime glass microspheres (Duke Standards, refractive index 1.5) of different diameters. The spheres filling fraction is 71\% and the nanoparticles concentration is 1\%. Due to the high refractive index contrast nanoparticles scatter light while glass spheres function only as spacers since they are index matched with the monomer. The diameters $\o_i$ of the spheres categories are as follows: 5, 8, 10, 15, 20, 30 ,40 ,50, 70, 100, 120, 140, 170, 200, 230, 280, 330, and 400 $\upmu$m. The power-law probability density to find a sphere of diameter $\o_i$ is obtained by distributing the number of spheres for each category as $N_i=C(\beta)\o_i^{-(\beta+1)}$, where $\beta$ can be arbitrary changed. The smaller $\beta$ the higher the probability of finding large spheres. The constant $C(\beta)$ depends on the total number of spheres in the samples. 
Samples are manufactured in a slab geometry by squeezing the mixture between two microscope glass slides (index matched with the monomer) until the sample thickness matches the largest sphere diameter.  Different thicknesses $L$ are obtained by removing all the spheres categories with a diameter larger than $L$. The amount of monomer and of nanoparticles grows with $L$, but the spheres filling fraction and the nanoparticles concentration are kept fixed. For each spheres category the diameters are distributed around the mean value $\o_i$ and since the thickness is determined by the few biggest values the measured $L$ can deviate from the nominal $\o_i$ by a 10$\%$. The reported $L$ do not contain the thickness of the slides. In the final stage samples are exposed to UV radiation for polymerization to ``freeze'' the components position.
\subsection{Homogeneous reference samples}
In the homogeneous reference samples microspheres are substituted with an equivalent volume of monomer, which leaves the final mixture  fluid. The slab geometry is obtained by first preparing a hosting cell of the desired $L$ with the same microscope slides used for the L\'evy glasses. The scattering monomer is inserted in the cell by means of infiltration. In the final stage samples are exposed to UV radiation for polymerization. 

\section{Experimental Setup}
\label{appendixB}

\begin{figure} [b!]
\centering 
\includegraphics [width=\columnwidth] {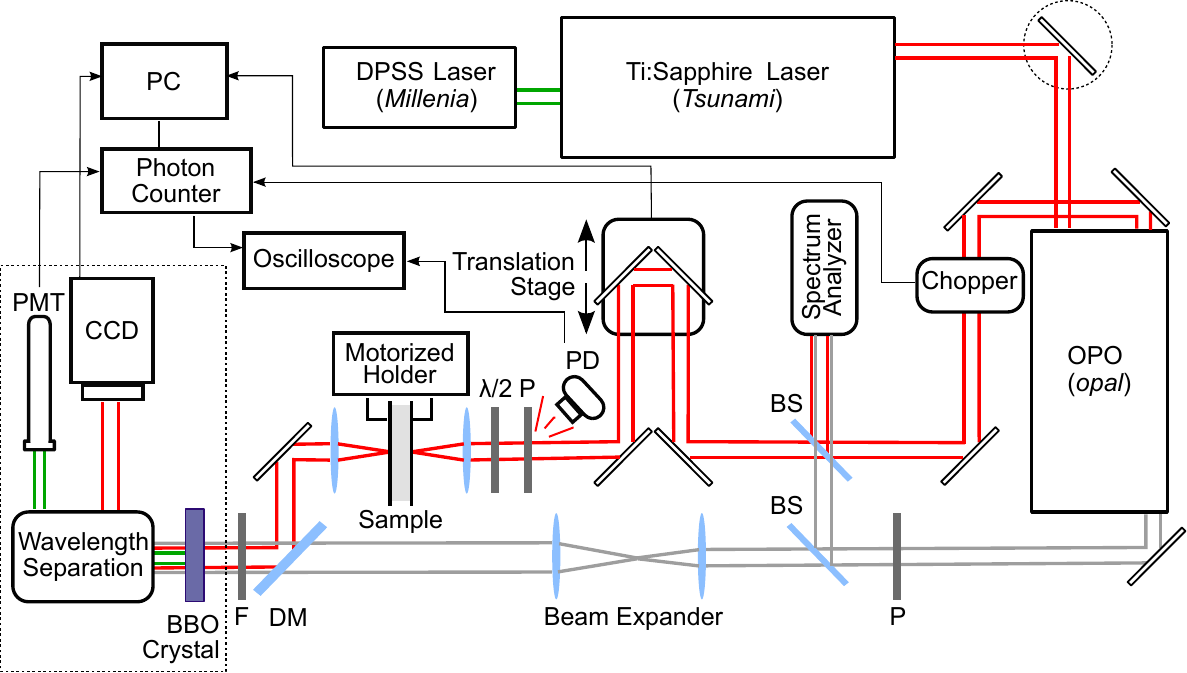} 
\caption{(Color online) Schematic of the experimental setup based on an optical gating technique: BS, beam splitter; P, polarizator; $\lambda/2$, half wave plate; DM, dicroic mirror; F, filter; PD, photodiode; PMT, photo multiplier tube. Dot contours indicate covering boxes.}
\label{fig:5}
\end{figure}

\subsection{Optical gating}
The setup (see Fig.~\ref{fig:5}) is based on an optical gating technique \cite{Shah:1988kl,svensson2013exploiting,Wiersma:1999kl} and involves the use of two ultrashort pulses ($\approx$130~fs) of different frequencies $\omega_1$ and $\omega_2$, whose wavelengths are respectively $\lambda_1$=810~nm  and $\lambda_2$=1540~nm. Pulses at $\omega_1$ are generated at a repetition rate of 82 MHz by a Ti:Sapphire mode-locked laser (\textsl{Tsunami}, Spectra Physics) pumped by a continuous-wave laser (\textsl{Millenia}, Spectra Physics). Pulses at $\omega_2$ are obtained by converting the pulse train at $\omega_1$ through an Optical Parametric Oscillator (\textsl{Opal}, Spectra Physics). The probe pulse at $\omega_1$ is incident onto the sample, interacts with its structure, and emerges with a stretched temporal profile $I_{\omega_{1}}(t)$. The gate pulse at $\omega_2$ travels undisturbed with its Gaussian temporal profile $I_{\omega_{2}}(t)$ on an independent line. The two optical signals are overlapped, in space and in time, into an optical non-linear crystal \textit{beta}-BBO (\textit{beta}-barium borate) which generates the sum-frequency  $I_{\omega_{1}+\omega_{2}}(t_d)\propto\int_0^{\infty}I_{\omega_{1}}(t)I_{\omega_{2}}(t-t_d)dt$, where $t_d$ is their relative time delay. Being the evolution of $I_{\omega_{1}}(t)$ much slower than that of $I_{\omega_{2}}(t)$, the latter can be approximated by a $\delta$-function so that the previous convolution becomes $I_{\omega_{1}+\omega_{2}}(t_d)\propto I_{\omega_{2}}I_{\omega_{1}}(t_d)$. By measuring the stationary signal $I_{\omega_{1}+\omega_{2}}(t_d)$ for different delays $t_d$ we retrieve the temporal profile $I_{\omega_{1}}(t)$ of the transmitted intensity. 
\subsection{Setup configuration}
The delay $t_d$ is changed by varying the optical path of the probe pulse through a motorized translation stage.
The sum-frequency is detected by a photomultiplier tube. Background noise suppression is obtained by chopping the beam (20~Hz) and by using  a Gated Integrator Photon Counter. A synchronization of the acquiring  temporal window is obtained by monitoring both the optical and the electronic trigger signals through an oscilloscope.
The injection and collection lenses are selected so as to image the whole transmitted profile on the BBO.
\subsection{Non linear conversion}
The phase-matching conditions on the BBO crystal are optimized for parallel collimated beams, implying that among the transmitted multiply scattered k-vectors only the forward directed ones are efficiently converted. The distribution of transmitted k-vectors is fully randomized (Lambertian distribution) after few scattering events  and it is time-independent. It follows that the time evolution of the measured intensity gives a correct representation of the time evolution of the total transmission since they are proportional at every instant through a time-independent constant. The correct functioning of the setup has been carefully tested on homogeneously disordered samples in a previous publication~\cite{svensson2013exploiting}. 

\section{Collapse on the intensity axis}
\label{appendixC}

It is theoretically predicted that the steady-state total transmission $T(L)= \int_0^\infty T(t,L)dt $ for annealed L\'evy walks with fractal dimension $d_\mathrm{w}<2$ through a slab of thickness $L$ scales as 
\begin{equation*}
T(L) \sim L^{-d_\mathrm{w}/2}, 
\end{equation*}
which is a generalized Ohm's law valid in the limit of very large $L$, where finite-size effects can be neglected~\cite{Davis1997,Buldyrev2001a}.

In this case the value of the exponent $\gamma$ in Eq.~(\ref{eq:scalingtransmission}) can be derived by rewriting Eq.~(\ref{eq:scalingtransmission}) as~\cite{Buonsante2011_PRE}
\begin{equation*}
T(t,L)=L^\gamma \tilde{T} \left (t/L^{d_\mathrm{w}} \right), 
\end{equation*}
and by imposing
\begin{equation*}
T(L)=\int_0^\infty T(t,L)dt=L^\gamma \int_0^\infty \tilde{T} \left(\frac{t}{L^{d_\mathrm{w}}} \right) \sim L^{-d_\mathrm{w}/2}.
\end{equation*}
By changing integration variable as $t\rightarrow t/L^{d_\mathrm{w}}$ the  previous similarity implies $\gamma=-\frac{3}{2}d_\mathrm{w}$.

We have applied this result to our case of finite-size samples by considering their effective thickness $L_\mathrm{eff}$ instead of the physical thickness $L$.

%

\end{document}